\RequirePackage[hyphens]{url}

\documentclass{ws-ijmpd}
\usepackage{amsmath,amssymb,graphicx,comment,listings,xcolor,paralist,hyperref}
\usepackage[super]{cite}

\allowdisplaybreaks

\definecolor{light-gray}{gray}{0.95}

\begin{document}

\title{Is graviton shot noise detectable?}
\author{Viktor T. Toth$^1$}
\address{$^1$Ottawa, K1N 9H5 ON Canada}


\maketitle

\begin{history}
\received{Day Month Year}
\revised{Day Month Year}
\accepted{Day Month Year}
\end{history}

\begin{abstract}
Direct detection of gravitons in gravitational experiments, including gravitational wave observatories, has been all but ruled out given the weak coupling between the gravitational field and matter. Here we propose an alternative: looking not for the presence but for the absence of graviton shot noise in gravitational wave data. Gravitational wave experiments detect very weak signals that correspond to a surprisingly small number of gravitons even at the relatively low frequencies that characterize signals from gravitational wave events. A detailed calculation, which also yields results that are consistent with the existing literature, demonstrates that graviton shot noise may be present at detectable levels in gravitational wave observations. The absence of elevated noise levels due to graviton shot noise, in turn, would indicate that gravity is not a quantum field theory with a conventional perturbative expansion at low energies.
\end{abstract}


\section{Introduction}

One of the greatest challenges that confront modern quantum theories of gravitation is detection. Gravity is exceedingly weak. Even if the entire Earth was used as a ``perfect'' graviton detector, it would not register more than roughly one atomic transition every {\em billion} years, mainly from the capture of solar thermal gravitons \cite{Dyson2012}. Such estimates seem to place any possible attempt to detect gravitons firmly beyond reach, in the realm of science-fiction, not practical science.

Nonetheless, it might be possible to detect gravitons. Or, at the very least, establish stringent limits that, while not necessarily confirming the existence of gravitons, can potentially {\em rule out} quantum theories of gravity by a failure to detect a key signature of gravitons.

As we shall see, conceivably the data are already in our hands. The detection of numerous gravitational wave events by the Laser Interferometer Gravitational-Wave Observatory (LIGO) experiment, specifically very ``clean'' signals from binary black hole and neutron star mergers, may be well-suited to check for the presence of a specific type of noise: graviton shot noise.

Professional low-light photographers are familiar with the problem of shot noise. When individual sensors, such as the pixels in a digital camera, are weakly illuminated, the number of photons captured by the sensor during the exposure time may be limited. This introduces the well-known Poisson-type shot noise that can make low light photographs acquire an unwelcome grainy appearance.

Shot noise is a frequently discussed subject in the literature of gravitational wave detection. The shot noise that is investigated in those discussions is shot noise associated with the optical measurement signal. The detected signal can be very weak, as the laser beam used in the interferometer will have covered the length of a multi-kilometer detector arm a great number of times, therefore, shot noise is of concern when interferometery is performed.

The existence of shot noise is independent of the specifics of the quantum theory of light. It is a consequence of quantization itself, notably that the amount of energy absorbed by the sensor is deposited in discrete units, or quanta. Individual photon energies are of course determined by the well-known expression $E_\gamma = \hbar\omega$ where $\hbar$ is the reduced Planck constant and $\omega$ is the angular frequency.

This same relationship applies to all massless particles, including hypothetical gravitons that, we expect, would appear as the quanta of gravitational radiation in the weak field, low energy limit, assuming of course that gravitation is governed by a quantum field theory in the first place. And while actual details obviously differ, fundamentally a gravitational wave detector is no different from a single pixel in the low-light photographer's camera: it absorbs signal energy one quantum at a time.

Therefore, if gravity is indeed quantized, shot noise will be present. This raises an obvious question: What is the magnitude of graviton shot noise in LIGO detections of gravitational wave events? Is this shot noise detectable?

In the following, we shall argue that indeed, shot noise is necessarily present in the gravitational wave signal, and that moreover, its detection is within the capabilities of existing gravitational wave observatories.

Of course, the presence of Poisson noise does not confirm quantum gravity. However, the absence of shot noise may be viewed as powerful evidence \textit{against} a quantum theory of gravitation. It is in light of this possibility that, in the following, we evaluate the amount of energy that an instrument or medium absorbs when it interacts with a gravitational wave signal.

We begin our investigation in Section~\ref{sec:opacity} by attempting to calculate the opacity of a medium to a gravitational wave that passes through it. Our calculation is purely classical, yet, as we shall see, it yields a result that is in close agreement with Dyson's famous estimate of the rate of graviton detection. Next, in Section~\ref{sec:shotnoise} we apply our approach to estimate the graviton shot noise that may be observed in a gravitational wave experiment. We derive the corresponding signal-to-noise ratio and compare it against other major known noise sources specific to the LIGO experiment. Finally, we offer our conclusions in Section~\ref{sec:summary}.

\section{Gravitational wave opacity}
\label{sec:opacity}

How can we calculate the gravitational wave opacity of a medium? Let us assume that
\begin{inparaenum}[a)]
\item the medium is homogeneous, and that
\item oscillations of the medium are forced,
i.e., the gravitational wave frequency is not a resonant frequency of the medium.
\end{inparaenum}
This calculation will inform us of the amount of power that the medium absorbs from the passing wave, and will also be directly useful when we later estimate the amount of signal power intercepted by a gravitational wave detector.

When a mass $m$ oscillates with angular frequency $\omega$ and amplitude $\tfrac{1}{2}\Delta L$, the associated kinetic energy is given by
\begin{align}
E = \tfrac{1}{2}m\omega^2 (\tfrac{1}{2}\Delta L)^2.\label{eq:oscE}
\end{align}
When the oscillation is forced, this energy must be replenished each half cycle. The corresponding power is given by
\begin{align}
P = \frac{1}{\pi}{\omega} E = \frac{1}{8\pi}m\omega^3\Delta L^2.\label{eq:oscP}
\end{align}
The displacement amplitude $\Delta L$ is related to the gravitational wave strain by $h = \Delta L/L.$ But what is $L$? In the case of an interferometer like LIGO, we would use the interferometer arm length as $L$, since the observed strain is measured as a change of the interferometer's length. However, when it comes to an extended medium, the effective $L$ may be larger; but no larger than the gravitational wave's wavelength, $L=\lambda=2\pi c/\omega$, thus
\begin{align}
P = \frac{1}{8\pi}m\omega^3\left(\frac{2\pi ch}{\omega}\right)^2=\tfrac{1}{2}\pi c^2h^2\omega m.
\end{align}

For a given density $\rho=m/V$, the absorption per unit volume is $\Pi=P/V$, or
\begin{align}
\Pi=\tfrac{1}{2} \pi c^2 h^2\omega\rho.
\end{align}

In contrast, the flux of a gravitational wave producing strain $h$ at angular frequency $\omega$ is given by \cite{LL1975}:
\begin{align}
\Phi = \frac{c^3\omega^2h^2}{16\pi G}.
\end{align}
This allows us to compute the absorption ratio,
\begin{align}
{\cal R}=\frac{\Pi}{\Phi} = \frac { 8\pi^2 G\rho}{c\omega},
\end{align}
or alternately, expressed as a cross-section per unit mass,
\begin{align}
\sigma = \frac { 8\pi^2 G}{c\omega}.\label{eq:sigma}
\end{align}

How does this compare against Dyson's famous prediction? Dyson calculates the per electron cross-section for graviton-electron scattering as $4\pi^2 G\hbar/c^3$. To convert this into a value of cross-section per unit mass, we must divide by the electron mass $m_e$, and obtain
\begin{align}
\sigma_e=\frac{4\pi^2 G\hbar}{c^3m_e}.
\end{align}
However, we must also consider that Dyson's focus was on calculating the number of gravitons that would be detected, even though in his calculation, individual thermal gravitons---emitted mostly by the deep solar interior, at a temperature of $T_{\odot}\sim 1.5\times 10^7$~K---had energies of $\sim 1.5$~keV, far in excess the activation energy of ground state electrons in a typical atom, $\sim 7.5$~eV. The excess energy is also transferred to the electron as kinetic energy, and ultimately absorbed by the medium by thermal scattering. This additional factor of $\alpha\sim 200$ must also be considered, leading us to the revised estimate of
\begin{align}
\sigma'_e=\alpha\frac{4\pi^2 G\hbar}{c^3m_e},
\end{align}
which would correctly capture the power absorbed by a medium as a gravitational wave passes through it.

To compare it against our estimate, we first recall the expression for the logarithmic peak of the Planckian radiation curve:

\begin{align}
\omega_{\rm logpeak}(T) = \frac{W(-4e^{-4})+4}{\hbar}k_BT,
\end{align}
where $W(z)$ is Lambert's $W$-function, satisfying the equation $W(z)e^{W(z)}=z$.

Comparing Dyson's result against our estimate, using $\omega=\omega_{\rm logpeak}(T_\odot)$, we find
\begin{align}
\frac{\sigma}{\sigma'_e}=\frac { 2 c^2m_e}{\alpha\omega_{\rm logpeak} \hbar}
=\frac { 2 c^2m_e}{\alpha(W(-4e^{-4})+4)k_B T_\odot}\sim 1.01.\label{eq:Dyson}
\end{align}
Encouragingly, this value indicates that our present approach is in near-perfect agreement with Dyson's result.

Without reading too much into this numerical agreement between the two calculations, it seems safe to conclude at the very least that it validates the current approach as a means to correctly model, to the right order of magnitude, the rate at which {\em the bulk} of a medium absorbs gravitons from a passing gravitational wave by any mechanism, not just electronic transitions.

\section{Graviton shot noise and LIGO}
\label{sec:shotnoise}

Shot noise is directly related to the quantization of a signal and, in particular, the number of particles that are captured by a detector. The magnitude of shot noise is characterized by the corresponding signal-to-noise ratio, which is $\sqrt{N}$, where $N$ is the number of particles that a detector detects instead of a continuous (classical) signal. So we have a very simple exercise: Count the number of gravitons that LIGO receives for the signal it measures.

Therefore, if we can estimate the number of gravitons that were captured by the LIGO detectors during a gravitational wave event, we can offer a sensible estimate of the shot noise that would be present in the measurement if gravity is indeed quantized.


One may argue that LIGO is not a conventional detector: an interferometer, not a ``graviton calorimeter''. However, this argument concerns strictly the interpretation of the measurement. What LIGO actually observes are changes in the relative signal paths along the two perpendicular arms towards the respective test masses. These test masses behave in analogy with Feynman's celebrated ``sticky beads'' (as retold by Bondi; see exercise 18.5 in Ref.~\citen{MTW}; also the foreword by Preskill and Thorne to Ref.~\citen{Feynman1995}) as they are brought into motion by the passing gravitational wave, moving as forced oscillators. The corresponding kinetic energy, transferred between the passing gravitational wave to the test masses, can be estimated.


Thus, we can view these detectors as the gravitational equivalent of a ``single-pixel'' sensor. That is, the detector effectively measures, during each sampling interval, the cumulative energy transferred to the detector test masses. To calculate the signal energy per sample, we begin by noting that LIGO has a $m = 40$~kg test mass at the end of a $L = 4$~km long arm. While the LIGO detection apparatus can be viewed as a mechanical oscillator, its resonance frequencies are in the $\sim 1$~Hz range, differing from typical detection frequencies by roughly two orders of the magnitude. Therefore, in the detection frequency range, LIGO may be viewed as a forced oscillator.

Earlier, we calculated the energy of an oscillator (\ref{eq:oscE}) and the amount of power required to maintain a forced oscillator (\ref{eq:oscP}). Dividing $P$ by the sampling rate $r$, we obtain the energy per sample, which we can compare against the energy of a single graviton, same as that of a single photon, at angular frequency $\omega$: $E_g = \hbar \omega$. We then calculate the number of gravitons per measurement sample:

\begin{align}
    N = \frac{P}{r\,E_g} = \frac{m\omega^3 \Delta L^2}{8\pi r\hbar\omega} = \frac{m\omega^2 h^2L^2}{8\pi \hbar r}.
\end{align}

For $m = 40$~kg, $\omega = 2\pi(250~{\rm Hz})$ (the peak frequency of, e.g., the GW150914 event \cite{LIGO2016}), $\Delta L = 10^{-21}\times 4$~km, $r = 4096$~Hz, we get
\begin{align}
N \approx 145.5\label{eq:N1}
\end{align}
gravitons per sample.

This implies ${\rm SNR} \approx 12.06$, or, in the language preferred by signal engineers, $\sim 10.8$~dB.

Can this result be valid? We have, after all, another expression for the gravitational wave cross-section per unit mass, in the form of Eq.~(\ref{eq:sigma}). However, this form would assume a gravitational wave antenna that operates with 100\% efficiency. Clearly, LIGO is no such antenna, as its physical size is more than two orders of magnitude smaller than the wavelength of the detected signal. In this case, we can characterize the antenna efficiency in the small antenna limit, $L\ll\lambda$, by \cite{Balanis1997}:
\begin{align}
\epsilon\approx \sin^2\frac{kL}{2}=\sin^2\frac{\pi L}{\lambda},
\end{align}
an expression that follows directly from assuming a pure sine wave signal amplitude with wavenumber $k=2\pi/\lambda$, and the resulting signal power distribution along the antenna length. Let us attempt to use this expression in the context of the GW150914 event, which took place $d\approx 1.4$ billion light years from Earth, and emitted the equivalent of approximately $5M_\odot$ (5 solar masses) of energy over the course of $\delta t\sim 200$~ms at a characteristic frequency of $\sim 250$~Hz. From this, we can easily compute the total number of gravitons: $N_g\approx 5M_\odot c^2/\hbar\omega\simeq 5.4\times 10^{78}$, which translates into $N_g/(4\pi d^2)\sim 2.46\times 10^{27}/{\rm m}^2$ gravitons on the surface of a sphere with a radius of $d=1.4\times 10^9$~ly over a duration of $0.2$~s. From this, we calculate
\begin{align}
N_0=\frac{\sigma m}{r\delta t}\frac{N_g}{4\pi d^2}\approx 1.35\times 10^6.
\end{align}
This would be the number of gravitons captured from the event by a $m=40$~kg gravitational wave antenna operating at 100\% efficiency. However, when we take the estimated antenna efficiency into account, we get instead
\begin{align}
N=\epsilon N_0\approx 146.2.\label{eq:N2}
\end{align}
Again, we see uncanny agreement between two independently calculated numbers, given by (\ref{eq:N1}) and (\ref{eq:N2}). This is a strong indication that our calculation, though approximate, reflects reality.

How can we make use of this result? The mere presence of Gaussian noise in the LIGO signal does not by itself imply that gravity is quantized. However, should we find that the noise is absent, or rather, that the observed noise is inconsistent with the presence of a $-10.8$~dB noise contribution by gravity quantization, this could prove to be a powerful negative result, strong evidence {\em against} quantum gravity.

Easier said than done! An SNR of 10.8~dB is a very respectable value, allowing for a clear signal to be discerned in he absence of other noise sources. Of course in the actual LIGO experiment, plenty of other noise sources are, in fact, present. Keeping noise under control has been one of the daunting challenges of the experiment.

To compare against other noise sources, we need to consider the spectral noise density associated with this result. For this, we first need to compute the strain associated with graviton shot noise:
\begin{align}
h_{\rm shot}=\frac{h}{\sqrt{N}}=\sqrt{\frac{8\pi\hbar r}{m}}\,\frac{1}{\omega L}.
\end{align}
Somewhat surprisingly, as the shot noise itself, $\sqrt{N}$ is a linear function of the strain $h$, the associated strain $h_{\rm shot}$ does not depend on the signal amplitude, only its frequency.

\begin{figure}
\centering\includegraphics[scale=0.5]{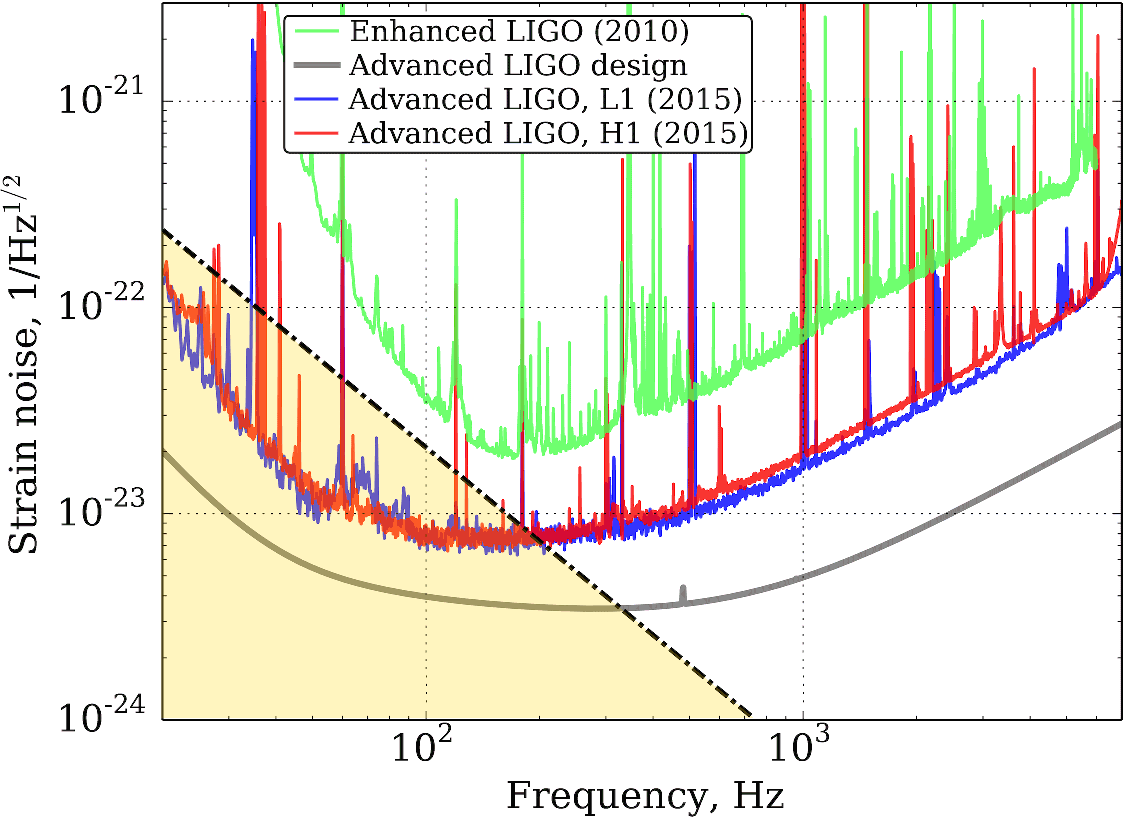}
\caption{\label{fig:strain}Sensitivity curve of the LIGO 
detector (adapted from Ref.~\citen{LIGO2016sens})
with the graviton shot noise (lower left shaded region demarcated by dash-dotted line) given by (\ref{eq:theNoise}) superimposed. 
}
\end{figure}

The corresponding noise spectral density is then
\begin{align}
S_{\rm grav}&{}=\sqrt{\frac{2\pi}{\omega}}h_{\rm shot}=\frac{4\pi}{L}\sqrt{\frac{\hbar r}{m\omega^3}}\nonumber\\
&{}\approx 5.2\times 10^{-24}\cdot\left(\frac{250~{\rm Hz}}{\omega}\right)^{3/2}~{\rm Hz}^{-1/2}.\label{eq:theNoise}
\end{align}
This value is, in fact, comparable to the known noise floor of the LIGO experiment at its peak sensitivity (Fig.~\ref{fig:strain}).

We may expect this noise to be independent of, and not correlated with, other noise sources in the LIGO experiment. Therefore, we expect that its contribution to the total spectral noise density $S_{\rm total}$ is characterized by the standard root-sum-square rule, i.e.,
\begin{align}
S_{\rm total}=\sqrt{S_{\rm grav}^2+S_{\rm nongrav}^2},
\end{align}
where we used $S_{\rm nongrav}$ to represent the noise spectral density associated with all nongravitational sources.

\section{Conclusions}
\label{sec:summary}

In the preceding section, we calculated the number of gravitons that correspond to the LIGO detection of major gravitational wave events, notably GW150914. This number may appear surprisingly low, especially considering the low frequency of the gravitational wave signal. Yet it is consistent with the exceptional sensitivity of LIGO and the weakness of a gravitational wave signal that arrives from cosmological distances. Moreover, we obtain consistent results using two independent methods of calculation (relying either on the measured strain or the estimated LIGO gravitational antenna efficiency.) Additionally, we have shown that our approach also yields results consistent with Dyson's celebrated lecture, demonstrating the validity of our semiclassical approach across different domains.


The surprisingly accurate numerical agreements shown in (\ref{eq:Dyson}) and also (\ref{eq:N1}) vs. (\ref{eq:N2}) are almost certainly coincidental, but the fact that three independent calculations yielded very similar results boosts our confidence that these results are valid, despite their apparent simplicity.

One important consequence is that such a small number of gravitons per sample may imply detectable graviton shot noise.

Graviton noise was previously considered by other authors. For instance, in [\citen{Wilczek2020,Wilczek2021a}], thermal graviton noise is investigated, but not shot noise. The same authors did consider ``quantum effects akin to graviton shot noise'' in another study\cite{Wilczek2021b} and asserted that their result falsifies an earlier claim \cite{Lieu2018} that such shot noise should be detectable. However, that earlier claim is based on the assumption that ``the GW does not actually exchange energy with the test mass''. In contrast, our
approach made it possible to treat the detector as a forced oscillator, consistent with the fact that the frequency of the detected gravitational wave is very different from the resonant frequency of the LIGO detector. As the detector's oscillation will be forced, detection requires a continuous exchange of energy between the gravitational wave and the detector.

Meanwhile, the authors of [\citen{Carney2024}] acknowledge the possibility that graviton shot noise may be present, and reach conclusions similar to what we demonstrate through explicit calculation: that ``simply observing gravitational shot noise [...] is not enough to demonstrate quantization.'' Recognition of this fact is precisely what led us to propose our current interpretation: Namely that, rather than treating the presence of shot noise as evidence of gravitons, we propose that persistent failure to detect graviton shot noise may falsify quantization.

As mentioned earlier, shot noise is Poisson noise. Given the anticipated number of gravitons per sample, it is indistinguishable from Gaussian noise. The mere presence of Gaussian noise in a LIGO measurement, however, cannot be considered as reliable evidence for quantum gravity. Gaussian noise can have many sources: interferometer shot noise, thermal noise, noise from residual gas pressure, technical noise sources are just some of the examples that can contribute to the LIGO noise budget. Therefore, even if higher-than-anticipated noise levels are detected at certain frequencies, the presence of such noise proves nothing.

On the other hand, what if the observed detector noise remains low? If,
say, in the frequency range $\sim$100~Hz, the overall observed strain noise remains below $\sim 10^{-23}$~Hz$^{-1/2}$, this would be powerful evidence {\em against} the presence of gravitons. This, then, is our main conclusion: the absence of noise consistent with Eq.~(\ref{eq:theNoise}), or with the shaded region of Fig.~(\ref{fig:strain}), would indicate, at the very least, that gravity cannot be modeled by a conventional quantum field theory that would yield enumerable excitations, i.e., gravitons, in the low energy perturbative limit.

In light of the anticipated magnitude of graviton shot noise, such a non-detection of gravitons seems eminently feasible. This powerful outcome would potentially falsify many families of quantum gravity theories.

\section*{Acknowledgments}

VTT thanks John W. Moffat for valuable comments and acknowledges the generous support of Plamen Vasilev and other Patreon patrons.

\bibliographystyle{ws-ijmpd}
\bibliography{refs}

\end{document}